\documentstyle[11pt,aaspp]{article}

\tighten
\input psfig.tex

\def\check_mode#1{\ifmmode{#1}\else{$#1$}\fi}

\def\lsim   {\check_mode{_<\atop^{\sim}}}

\slugcomment{COBE Preprint \# 96-04, Submitted to {\it ApJ Letters}}
\begin{document}

\title{Band Power Spectra in the {\it COBE}\altaffilmark{1} DMR 4-Year 
Anisotropy Maps}

\author{ G. Hinshaw\altaffilmark{2,3},
         A.J. Banday\altaffilmark{2,4},
         C.L. Bennett\altaffilmark{5},
         K.M. G\'orski\altaffilmark{2,6},
         A. Kogut\altaffilmark{2}, \nl
         G.F. Smoot\altaffilmark{7,8}
         \& E.L. Wright\altaffilmark{9}}

\altaffiltext{1}{The National Aeronautics and Space Administration/Goddard 
Space Flight Center (NASA/GSFC) is responsible for the design, development, 
and operation of the Cosmic Background Explorer ({\it COBE}).
Scientific guidance is provided by the {\it COBE} Science Working Group.
GSFC is also responsible for the development of the analysis software and
for the production of the mission data sets.}
\altaffiltext{2}{Hughes STX Corporation, 
                 Laboratory for Astronomy and Solar Physics,
                 Code 685, NASA/GSFC, 
                 Greenbelt MD 20771.}
\altaffiltext{3}{e-mail: hinshaw@stars.gsfc.nasa.gov}
\altaffiltext{4}{Current address:
                 Max Planck Institut fur Astrophysik,
                 85740 Garching Bei Munchen
                 Germany.}
\altaffiltext{5}{Laboratory for Astronomy and Solar Physics,
                 Code 685, NASA/GSFC, 
                 Greenbelt MD 20771.}
\altaffiltext{6}{On leave from Warsaw University Observatory, 
                 Aleje Ujazdowskie 4, 00-478 Warszawa, Poland.}
\altaffiltext{7}{Department of Physics, U.C. Berkeley, 
                 Berkeley, CA 94720.}
\altaffiltext{8}{Lawrence Berkeley Laboratory, Bldg 50-351, 
                 University of California, Berkeley CA 94720.}
\altaffiltext{9}{UCLA Astronomy, P.O. Box 951562, Los Angeles CA 90095-1562.}

%\altaffiltext{8}{UCSB Physics Department, Santa Barbara CA 93106.}

\begin{abstract}
We employ a pixel-based likelihood technique to estimate the angular power 
spectrum of the {\it COBE} Differential Microwave Radiometer (DMR) 4-year 
sky maps.  The spectrum is consistent with a scale-invariant power-law form 
with a normalization, expressed in terms of the expected quadrupole anisotropy, 
of $Q_{rms-PS \vert n=1} = 18 \pm 1.4$ $\mu$K, and a best-fit spectral index 
of 1.2 $\pm$ 0.3.  The normalization is somewhat smaller than we concluded 
from the 2-year data, mainly due to additional Galactic modeling.  We extend 
the analysis to investigate the extent to which the ``small" quadrupole 
observed in our sky is statistically consistent with a power-law spectrum.  
The most likely quadrupole amplitude is somewhat dependent on the details of 
Galactic foreground subtraction and data selection, ranging between 7 and 10 
$\mu$K, but in no case is there compelling evidence that the quadrupole is 
too small to be consistent with a power-law spectrum.  We conclude with a 
likelihood analysis of the band power amplitude in each of four spectral bands 
between $\ell =$ 2 and 40, and find no evidence for deviations from a simple 
power-law spectrum.
\end{abstract}

\keywords{cosmic microwave background --- cosmology: observations}

\clearpage

\section{Introduction}

The detection of large angular scale anisotropies in the Cosmic Microwave 
Background (CMB) radiation was first reported by the {\it COBE}-DMR experiment
in 1992 (Smoot et al. 1992; Bennett et al. 1992; Wright et al. 1992; Kogut et 
al. 1992).  The initial detection was based only on the first year of flight 
data.  Since that time the DMR Team processed and analyzed the first two 
years of data and found results to be consistent with the first year 
results (Bennett et al. 1994, G\'orski et al. 1994, Wright et al. 1994a).  We 
have now processed and analyzed the full 4-years of DMR observations: this 
paper is one of a series describing the results of our analysis.  The maps and 
an overview of the scientific results are given in Bennett et al. (1996).

In this paper we analyze the angular power spectrum of the 4-year DMR maps 
using a pixel based likelihood technique which was first applied to the 2-year 
data by Tegmark \& Bunn (1995).  We extend previous work by considering 
several parameterization of the angular power spectrum.  The simplest model 
for large angular scale anisotropy is the power-law model parameterized by a 
normalization, $Q_{rms-PS}$, and spectral index, $n$.  It is of interest to 
separate the quadrupole anisotropy from the rest of the power spectrum since 
it is most plausibly contaminated by Galactic emission, and, in some models, 
is predicted to deviate from the simple power-law form.  We extend our 
likelihood analysis to a three parameter model in which the quadrupole is fit 
independent of the higher-order power (which is assumed to follow a power-law, 
see \S 4 for details).  Lastly we consider band power estimates in which the anisotropy is assumed to 
be scale-invariant in each of four modestly narrow $\ell$ bands, chosen to 
have roughly comparable sensitivity.  The results are compared to the power-law
fits and indicate that the anisotropy has no significant deviation from a 
power-law form.

\section{Data Selection}

The DMR experiment has produced two independent microwave maps (A and B) at 
each of 3 frequencies (31.5, 53 and 90 GHz).  The results presented here are 
based on linear combinations of all 6 channel maps.  The combination 
coefficients are dictated by the sensitivities of the individual channels and 
on considerations of Galactic foreground removal.  Kogut et al. (1996a,1996b) 
have revisited the question of Galactic emission in the 2- and 4-year DMR data 
at high latitudes, and conclude that there is statistically significant 
evidence for a weak Galactic signal at all three frequencies, even at 
latitudes $|b| \geq 20^{\circ}$.  Thus, for the 4-year analysis, we take a 
more aggressive approach to Galactic foreground removal than we have 
previously: first, we extend the Galactic plane cut of $20^{\circ}$ with 
additional cuts, guided by the $COBE$-DIRBE 140 $\mu$m map (Bennett et al. 
1996).  The number of pixels surviving the cut is 3881 in Galactic 
coordinates.  Second, we model and remove residual high-latitude Galactic 
emission in two complementary ways, described below.  In all, we analyze three 
separate maps in this paper: the first map is a weighted average of all 
six DMR channel maps with no residual Galaxy emission subtracted.  We denote 
this map ``31+53+90".  The second is the same weighted average map as the first 
with best-fit Galaxy template maps subtracted from each channel prior to 
averaging (Kogut et al. 1996b).  We denote this the ``Correlation" model map.  
The third is a linear combination of all six channels with coefficients 
designed to maximize sensitivity subject to the constraint that any signal 
with a free-free frequency spectrum ($\beta_{ff} = -2.15$) cancels.  This map 
also has best-fit synchrotron and dust emission templates subtracted prior to 
averaging (Kogut et al. 1996b).  We denote this the ``Combination" model map.  
The specific coefficients used to construct these maps are given in Table 1 of 
Hinshaw et al. (1996).  In all the analyses below we use the maps pixelized in 
Galactic coordinates.  G\'orski et al. (1996) and Banday et al. (1996) have 
analyzed both the Galactic and ecliptic maps in detail.  Since our results 
for the Galactic maps are in agreement with theirs, where comparable, we defer 
to those papers for a comparison of the Galactic and ecliptic maps.

\section{Method}

Most cosmological models make predictions for the mean angular power spectrum 
of CMB anisotropies, the coefficients, $C_{\ell}$.  For a rotationally 
invariant theory, the $C_{\ell}$ specify the expected variance in each 
spherical harmonic mode in a Fourier expansion of the sky temperature
$T(\theta,\phi) = \sum_{\ell,m} a_{\ell m} Y_{\ell m}(\theta,\phi)$
with $\langle a_{\ell m}a^{*}_{\ell' m'} \rangle 
= C_{\ell}\,\delta_{\ell \ell'}\,\delta_{mm'}$.  For a given power spectrum, 
$C_{\ell}$, the implied covariance between map pixels $i$ and $j$ is given by
\begin{equation}
M_{ij} = \langle T_i T_j \rangle = \frac{1}{4\pi} \sum_{\ell} \, (2\ell+1) 
\, W^{2}_{\ell} \, C_{\ell} \, P_l(\hat{n_i} \cdot \hat{n_j})
\label{cov_eq}
\end{equation}
where $T_i$ is the temperature in pixel $i$ of a map, the angled brackets 
denote a universal ensemble average, $W^{2}_{\ell}$ is the experimental window 
function that includes the effects of beam smoothing and finite pixel size, 
$C_{\ell}$ is the power spectrum, $P_l(\hat{n_i} \cdot \hat{n_j})$ is the 
Legendre polynomial of order $\ell$, and $\hat{n_i}$ is the unit vector 
towards the center of pixel $i$.  For Gaussian fluctuations, the covariance 
matrix fully specifies the statistics of the temperature fluctuations.  The 
probability of observing a map with pixel temperatures $\vec{T}$, given a 
power spectrum $C_{\ell}$, is
\begin{equation}
P(\vec{T} \vert C_{\ell})\,d\vec{T} = \frac{d\vec{T}}{(2\pi)^{N/2}} \:
\frac{e^{-\frac{1}{2}\vec{T}^{T}\cdot M(C_{\ell})^{-1} \cdot \vec{T}}}
     {\sqrt{\det M(C_{\ell})}}
\label{prob_eq}
\end{equation}
where $N$ is the number of pixels in the map.  Assuming a uniform prior 
distribution of cosmological model parameters, the probability of a power 
spectrum $C_{\ell}$, given a map $\vec{T}$, is then
\begin{equation}
{\cal L}(C_{\ell} \vert \vec{T}) \propto 
\frac{e^{-\frac{1}{2}\vec{T}^{T}\cdot M^{-1}(C_{\ell}) \cdot \vec{T}}}
     {\sqrt{\det M(C_{\ell})}}.
\label{like_eq}
\end{equation}

In the following section, we evaluate the above likelihood function using 
three different parameterizations of the power spectrum, $C_{\ell}$.  To test 
the effects of data selection and Galaxy modeling, we analyze three separate 
DMR maps, as specified in \S 2.  To make the analysis computationally 
efficient, we have degraded the maps by one step in pixel resolution 
(to index level 5) for which there are 1536 pixels in the full sky, and 954 
pixels surviving the extended Galaxy cut.  We account for the effects of 
smoothing due to pixelization by including a term in the window function:
$W_{\ell} = G_{\ell} F_{\ell}$.  The $G_{\ell}$ are the Legendre coefficients 
of the DMR beam pattern, tabulated by Wright et al. (1994b).  The $F_{\ell}$ 
are the Legendre coefficients for a circular top-hat function with area equal 
to the pixel area.  The coefficients for index level 5 pixels are available on 
request.

We ignore the contribution of the monopole and dipole moments in the maps 
since these modes are either unconstrained by the data (the monopole), 
or are dominated by local effects (the dipole).  In principle, this is achieved 
by integrating the likelihood over the modes $C_{0}$ and $C_{1}$; in practice, 
we set these terms to large positive constants when evaluating the covariance 
matrix.  We have found that setting $C_{0} = C_{1} = 10^{8}$ $\mu$K$^{2}$ 
renders the likelihood insensitive to monopole and dipole moments of several 
hundred $\mu$K, without compromising the inversion of the covariance matrices. 
(Note that the analyzed maps have approximately zero mean, by construction, 
and approximately zero dipole since an estimate of the CMB dipole is removed 
during the raw data processing.)  We assume the noise in the sky maps is 
uncorrelated from pixel to pixel (Lineweaver et al. 1994), which adds a 
diagonal contribution to the pixel covariance matrix in Equation \ref{cov_eq}.  
Tegmark \& Bunn (1995) have shown the assumption of uncorrelated noise to be 
an excellent approximation for this application.  The noise per pixel is 
derived from the noise per observation, given in Table 1 of Bennett et al. 
(1996), and the number of observations per pixel.

\section{Results}

We consider three parameterizations of the angular power spectrum.  First, 
we adopt the power-law model, parameterized by the amplitude of the mean 
quadrupole anisotropy, $Q_{rms-PS}$, and the power-law spectral index $n$.  
Specifically (Bond \& Efstathiou, 1987)
\begin{equation}
C_{\ell} = C_{\ell}(Q_{rms-PS},n) \equiv 
(4\pi/5)Q_{rms-PS}^{2} \, \frac{\Gamma(\ell+(n-1)/2)\Gamma((9-n)/2)}
                               {\Gamma(\ell+(5-n)/2)\Gamma((3+n)/2)}
\label{qn_model_eq}
\end{equation}
This model is extended to study the quadrupole anisotropy by parameterizing 
the power at $\ell$ = 2 separately
\begin{equation}
C_{\ell} =  \left\{ \begin{array}{ll}
                    C_2                    & \mbox{$\ell = 2$} \\
                    C_{\ell}(Q_{rms-PS},n) & \mbox{$\ell \geq 3$}
                    \end{array}
            \right.
\label{q2qn_model_eq}
\end{equation}
The most-likely value of $C_2$ that results from this model is closely related 
to the quadrupole anisotropy observed in our sky, which we denote $Q_{rms}$;  
the precise connection is discussed below.  Note also that the power-law 
parameters, $Q_{rms-PS}$ and $n$, inferred from this model are essentially 
the same as those derived from marginalizing over $C_2$ since they are only 
weakly coupled to $C_2$.  Lastly, we study a model in which the spectrum is 
taken to be scale-invariant in each of four relatively narrow $\ell$ bands, 
and let the amplitude in each be a free parameter
\begin{equation}
\ell(\ell+1)\,C_{\ell} =  \left\{ \begin{array}{ll}
             (24 \pi / 5)\, Q_{\alpha}^2 & \mbox{$2 \leq \ell \leq 5$} \\
             (24 \pi / 5)\, Q_{\beta}^2 & \mbox{$6 \leq \ell \leq 10$} \\
             (24 \pi / 5)\, Q_{\gamma}^2 & \mbox{$11 \leq \ell \leq 20$} \\
             (24 \pi / 5)\, Q_{\delta}^2 & \mbox{$21 \leq \ell \leq 40$}
             \end{array}
             \right.
\label{band_model_eq}
\end{equation}
Note that within each band, the amplitude parameters $Q_{\alpha\cdots\delta}$
correspond to $Q_{flat}$ as defined by Scott, Silk \& White (1995).  The 
spectral band widths were chosen to give roughly equal sensitivity in each 
band except the highest which suffers loss of signal due to the $7^{\circ}$ 
beam width.  The model is designed to probe for deviations from a power-law 
while maintaining computational feasibility.

The fits to power-law spectra, including the quadrupole in the analysis, are 
summarized in Table \ref{qn_table}.  The results are generally consistent with 
the 2-year data.  The overall normalization is slightly smaller due to the 
additional Galactic cutting and modeling.  The most-likely spectral index is 
slightly greater than unity, while the quadrupole normalization for a 
scale-invariant spectrum ranges from 17.2 to 18.4 $\mu$K, depending on 
Galactic model.  For comparison, the scale-invariant normalization derived 
from the weighted average map using a straight $20^{\circ}$ cut with no 
additional Galactic modeling is 20.1 $\mu$K, 1.6 $\mu$K higher than we obtain 
with the extended cut, and comparable to the normalization quoted by G\'orski 
et al. (1994) for the 2-year data (using a straight cut).  In assessing the 
results obtained from the three maps, we note that the DIRBE 140 $\mu$m map 
appears to trace the bulk of the free-free emission seen by DMR (Kogut et al. 
1996b), and since the Correlation map is more sensitive than the Combination 
map, we give it more weight in our conclusions.  Taken together, the results 
in Table 1 are consistent with a spectral index of 1.2 $\pm$ 0.3 and a 
scale-invariant quadrupole normalization of 18 $\pm$ 1.4 $\mu$K.

\begin{figure}[t]
\psfig{file=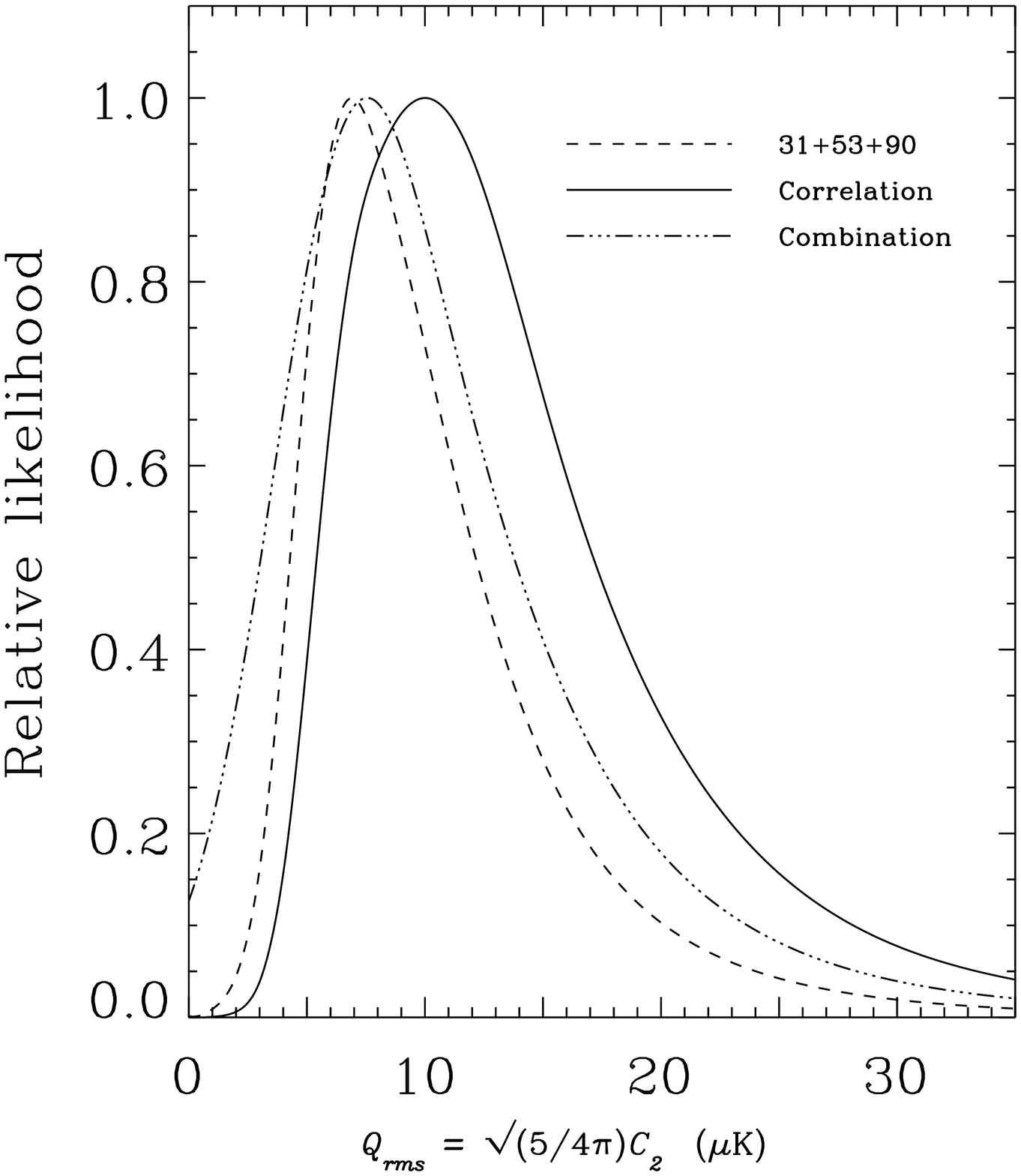,width=6.0in}
\caption{The likelihood function for the mean quadrupole moment observed in 
our sky for the three maps defined in \S 2.  The curves include cosmic 
variance and instrument noise.  In all cases the most likely quadrupole is 
smaller than that favored by the power-law fits to the full data, but the 
likelihoods are all sufficiently broad to encompass the case $Q_{rms} = 
Q_{rms-PS}$.  The effect of Galactic modeling on $Q_{rms}$ is relatively 
modest, but it does have a significant effect on the phase of the quadrupole, 
particularly the coefficient $a_{20}$ (Kogut et al. 1996)}
\label{q2_like_fig}
\end{figure}

The fits to power-law spectra with the quadrupole parameterized independently 
are summarized in Table \ref{q2qn_table}.  The first three columns summarize 
the power-law portion of the spectrum in the same format as Table 
\ref{qn_table}.  These results are based on slicing the 3-dimensional 
likelihood at the maximum likelihood value for $C_2$, but the results are 
only weakly dependent on $C_2$ and thus are effectively equivalent to standard 
power-law fits that ignore the quadrupole.  The last column of Table 
\ref{q2qn_table} gives the 68\% confidence interval for $C_2$ expressed in 
terms of $Q_{rms} \equiv \sqrt{(5/4 \pi) C_2}$.  The mode gives a 
self-consistent estimate of the quadrupole moment observed in our sky, while 
the confidence range accounts for both instrument noise {\it and} cosmic 
variance.  A complimentary approach to analyzing the quadrupole, based on 
fitting and squaring $a_{2m}$ coefficients (Kogut et al. 1996b) gives 
consistent results, after accounting for the bias introduced by uncertainties 
in the $a_{2m}$.  This approach demonstrates the importance of modeling the 
residual galactic foreground emission since it contributes significantly to 
the quadrupole emission.  As with previous analyses, the observed quadrupole 
is smaller than that expected from the power-law fits: the most-likely 
amplitude ranges from 6.9 to 10.0 $\mu$K depending on Galaxy model.  Figure 
\ref{q2_like_fig} shows the full likelihood for $Q_{rms}$ for each map 
analyzed.  It is important to stress that while the quadrupole in our sky is 
most likely $\sim$10 $\mu$K, the cosmic variance combined with experimental 
uncertainties are so large that its value is easily consistent with a 
power-law model of anisotropy.  For example, the likelihood for $Q_{rms}$ 
derived from the Correlation map implies there is a 22\% chance that 
$Q_{rms}$ exceeds 18 $\mu$K, the value favored in a scale-invariant power-law 
model.

\begin{figure}[t]
\psfig{file=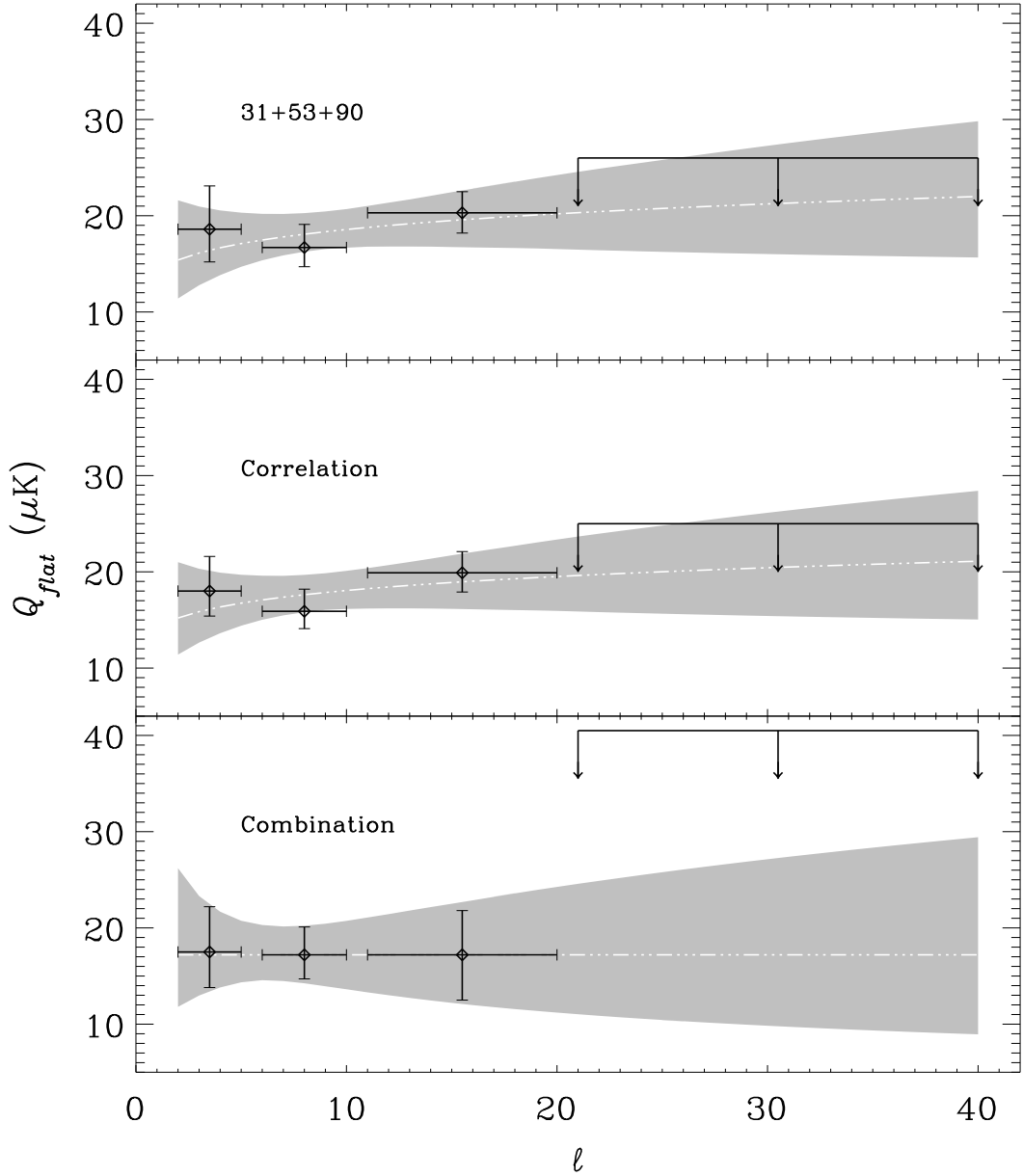,width=6.0in}
\caption{Power spectral estimates for three maps defined in \S 2.
The points with error bars give the most likely band power amplitude within 
each $\ell$ band, as indicated by the horizontal error bars, under the 
assumption that the power spectrum is scale-invariant within each band.  The 
fourth band, $21 \leq \ell \leq 40$, is plotted as a 95\% CL upper limit, 
since there is no significant detection of power in this band.  The vertical 
errors include both noise and cosmic variance.  The shaded region indicates 
the locus of the mean of acceptable power-law models, i.e., those models 
within the 68\% confidence region in the $(Q_{rms-PS},n)$ plane, as determined 
from the power-law fits to each map.  The dashed white line within the shaded 
region gives the mean power spectrum for the most-likely power-law model.  
Note that any given realization of a power-law spectrum will, in general, 
deviate from the mean spectra plotted here due to cosmic variance.}
\label{band_fig}
\end{figure}

The band power fits are summarized in Table \ref{band_table}, and are plotted 
in Figure \ref{band_fig}.  The vertical uncertainties in the figure indicate 
the extent of the 68\% confidence interval in each band power parameter when 
the other 3 are fixed at their maximum likelihood value.  These uncertainties 
include both instrument noise and cosmic variance.  The horizontal errors bars 
represent the extent of each band, as defined in Equation \ref{band_model_eq}.
The covariance between bands, which arises from the Galaxy cut and from 
non-uniform sky coverage, is quite small: roughly 10\% of the parameter 
variance for neighboring bands, and less for non-neighboring bands.  Note that 
each of the three lowest bands have consistently significant detections of 
power, while, in all cases, the highest band, from 21 to 40, does not.  Thus, 
we only plot 95\% confidence upper limits for this band.  To compare the band 
power fits to the power-law fits we have also plotted the 68\% confidence 
locus of acceptable power-law models in Figure \ref{band_fig}.  More precisely, 
the dashed white line in the figure is the mean power spectrum for the 
most-likely power-law model, while the grey band represents the locus of mean 
power spectra within the 68\% confidence region in the ($Q_{rms-PS},n$) plane.  
The general agreement between the power-law model and the band power model is 
an indication that there are no significant wide band deviations (with $\Delta 
\ell \sim$ a few) from a simple power-law in the low-$\ell$ anisotropy 
spectrum.  The band power amplitude in the highest $\ell$ band we probe 
is consistently low, but this estimate is rather sensitive to the details of 
the beam and pixelization filters, and to the level of noise in the maps, so
the uncertainty attached to this estimate is quite large.  To date, two other 
experiments have measured anisotropy on angular scales probed by the DMR: 
FIRS (Ganga et al. 1994) and Tenerife (Hancock et al. 1994).  Both experiments 
report significant detections of anisotropy: the FIRS team quotes $Q_{flat}$ = 
19 $\pm$ 5 $\mu$K for $\ell \lsim 30$, while the Tenerife team quotes 
$Q_{flat}$ = 26 $\pm$ 6 $\mu$K for $13 \lsim \ell \lsim 30$, both of which are 
consistent with DMR.

\section{Conclusions}

We have estimated various parameterizations of the angular power spectrum in
the {\it COBE}-DMR 4-year sky maps.  We find the results to be generally 
consistent with the first and second year results.  The data are consistent 
with a scale-invariant spectrum with a quadrupole normalization of 18 $\pm$ 
1.4 $\mu$K, and a best-fit spectral index of 1.2 $\pm$ 0.3.  The quadrupole 
anisotropy is somewhat smaller than the best-fit power-law spectrum would 
prefer, but the discrepancy is not statistically significant when we take 
account of Galactic modeling uncertainties, instrument noise, and cosmic 
variance.  We have further analyzed the spectrum in each of four $\ell$ bands 
and find no evidence for significant, wide-band deviations from a simple 
power-law form.

We gratefully acknowledge the many people who made this paper possible: the 
NASA Office of Space Sciences, the {\it COBE} flight operations team, and all 
of those who helped process and analyze the data.  We also thank Charley 
Lineweaver for useful comments on our manuscript.

\clearpage
\begin{planotable}{llll}
\tablewidth{4.5in}
\tablecaption{Power-law Spectral Parameters}
\tablehead{ \colhead{Map}                                      &
            \colhead{$n$\tablenotemark{a}}                     &
            \colhead{$Q_{rms-PS}$\tablenotemark{b}}            &
            \colhead{$Q_{rms-PS \vert n=1}$\tablenotemark{c}}  \nl
            \colhead{}                                         &
            \colhead{}                                         &
            \colhead{($\mu$K)}                                 &
            \colhead{($\mu$K)}                                 }
\startdata
31+53+90\tablenotemark{d}       & $1.25^{+0.26}_{-0.29}$ 
    & $15.4^{+3.9}_{-2.9}$ & $18.4^{+1.4}_{-1.3}$ \nl
Correlation\tablenotemark{d}    & $1.23^{+0.26}_{-0.27}$ 
    & $15.2^{+3.6}_{-2.8}$ & $17.8^{+1.3}_{-1.3}$ \nl
Combination\tablenotemark{d}    & $1.00^{+0.40}_{-0.43}$ 
    & $17.2^{+5.6}_{-4.0}$ & $17.2^{+1.9}_{-1.7}$ \nl
%31+53+90\tablenotemark{d}       & $1.25^{+0.26}_{-0.29}$ 
%    & $15.43^{+3.89}_{-2.85}$ & $18.43^{+1.35}_{-1.28}$ \nl
%Correlation\tablenotemark{d}    & $1.23^{+0.26}_{-0.27}$ 
%    & $15.24^{+3.56}_{-2.78}$ & $17.82^{+1.33}_{-1.25}$ \nl
%Combination\tablenotemark{d}    & $1.00^{+0.40}_{-0.43}$ 
%    & $17.19^{+5.56}_{-3.96}$ & $17.17^{+1.85}_{-1.70}$ \nl
\tablenotetext{a}{Mode and $\pm$68\% confidence interval for the projection of 
the 2-dimensional likelihood ${\cal L}(Q_{rms-PS},n)$ on to $n$.}
\tablenotetext{b}{Mode and $\pm$68\% confidence interval for the projection of 
the 2-dimensional likelihood ${\cal L}(Q_{rms-PS},n)$ on to $Q_{rms-PS}$.}
\tablenotetext{c}{Mode and $\pm$68\% confidence interval for the slice of 
the 2-dimensional likelihood ${\cal L}(Q_{rms-PS},n)$ at $n=1$.}
\tablenotetext{d}{Linear combination coefficients for the maps analyzed here 
are given in Table 1 of Hinshaw et al. (1996).}
\label{qn_table}
\end{planotable}

\begin{planotable}{lllll}
\tablewidth{4.5in}
\tablecaption{Quadrupole + Power-law Spectral Parameters}
\tablehead{ \colhead{Map}                                      &
            \colhead{$n$\tablenotemark{a}}                     &
            \colhead{$Q_{rms-PS}$\tablenotemark{b}}            &
            \colhead{$Q_{rms-PS \vert n=1}$\tablenotemark{c}}  &
            \colhead{$Q_{rms}$\tablenotemark{d}}               \nl
            \colhead{}                                         &
            \colhead{}                                         &
            \colhead{($\mu$K)}                                 &
            \colhead{($\mu$K)}                                 &
            \colhead{($\mu$K)}                                 }
\startdata
31+53+90\tablenotemark{e}      & $1.09^{+0.29}_{-0.30}$ &
 $17.5^{+4.7}_{-3.6}$ & $18.7^{+1.4}_{-1.3}$ & $6.9^{+5.4}_{-2.7}$ \nl
Correlation\tablenotemark{e}   & $1.09^{+0.29}_{-0.31}$ &
 $17.0^{+4.7}_{-3.6}$ & $18.1^{+1.4}_{-1.3}$ & $10.0^{+6.5}_{-4.4}$ \nl
Combination\tablenotemark{e}   & $0.57^{+0.44}_{-0.49}$ &
 $23.0^{+8.4}_{-5.7}$ & $17.9^{+1.9}_{-1.8}$ &  $7.6^{+6.2}_{-4.5}$ \nl
%31+53+90\tablenotemark{e}      & $1.09^{+0.29}_{-0.30}$ &
% $17.50^{+4.72}_{-3.60}$ & $18.65^{+1.37}_{-1.29}$ & $6.91^{+5.40}_{-2.67}$ \nl
%Correlation\tablenotemark{e}   & $1.09^{+0.29}_{-0.31}$ &
% $17.02^{+4.71}_{-3.58}$ & $18.11^{+1.37}_{-1.28}$ & $10.01^{+6.49}_{-4.41}$ \nl
%Combination\tablenotemark{e}   & $0.57^{+0.44}_{-0.49}$ &
% $22.98^{+8.39}_{-5.69}$ & $17.87^{+1.92}_{-1.77}$ &  $7.56^{+6.16}_{-4.48}$ \nl
\tablenotetext{a}{Mode and $\pm$68\% confidence interval for the projection of 
the 2-dimensional likelihood ${\cal L}(Q_{rms-PS},n)$ on to $n$.  
${\cal L}(Q_{rms-PS},n)$ is the 3-dimensional likelihood 
${\cal L}(Q_{rms},Q_{rms-PS},n)$ evaluated at the maximum likelihood value of 
$Q_{rms}$.}
\tablenotetext{b}{Mode and $\pm$68\% confidence interval for the projection of 
the 2-dimensional likelihood ${\cal L}(Q_{rms-PS},n)$ on to $Q_{rms-PS}$.}
\tablenotetext{c}{Mode and $\pm$68\% confidence interval for the slice of 
the 2-dimensional likelihood ${\cal L}(Q_{rms-PS},n)$ at $n=1$.}
\tablenotetext{d}{Mode and $\pm$68\% confidence interval for the slice of 
the 3-dimensional likelihood ${\cal L}(Q_{rms},Q_{rms-PS},n)$ at the maximum 
likelihood values of $Q_{rms-PS}$ and $n$.  Similar results are obtained by 
marginalizing over $Q_{rms-PS}$ and $n$.}
\tablenotetext{e}{Linear combination coefficients for the maps analyzed here 
are given in Table 1 of Hinshaw et al. (1996).}
\label{q2qn_table}
\end{planotable}

\begin{planotable}{lllll}
\tablewidth{4.5in}
\tablecaption{Band Power Spectral Parameters\tablenotemark{a}}
\tablehead{ \colhead{Map}                      &
            \multicolumn{4}{c}{-----Band-----} \nl
            \colhead{}                         &
            \colhead{$ 2 \leq \ell \leq  5$}   &
            \colhead{$ 6 \leq \ell \leq 10$}   &
            \colhead{$11 \leq \ell \leq 20$}   &
            \colhead{$21 \leq \ell \leq 40$}   }
\startdata
31+53+90\tablenotemark{b}      & $18.6^{+4.5}_{-3.4}$ &
 $16.7^{+2.4}_{-2.0}$ & $20.3^{+2.2}_{-2.1}$ & $1.0^{+13.2}_{-1.0}$ \nl
Correlation\tablenotemark{b}   & $18.0^{+3.6}_{-2.6}$ &
 $15.9^{+2.3}_{-1.8}$ & $19.9^{+2.2}_{-2.0}$ & $0.8^{+12.6}_{-0.8}$ \nl
Combination\tablenotemark{b}   & $17.5^{+4.7}_{-3.7}$ &
 $17.2^{+2.9}_{-2.5}$ & $17.2^{+4.6}_{-4.7}$ & $0.1^{+22.2}_{-0.1}$ \nl
\tablenotetext{a}{Mode and $\pm$68\% confidence interval for the band power 
amplitudes, expressed in terms of $Q_{flat}$.  $Q_{flat}$ is the quadrupole 
normalization expected for a scale-invariant power-law spectrum within the 
specified range of $\ell$.  The units are $\mu$K.}
\tablenotetext{b}{Linear combination coefficients for the maps analyzed here 
are given in Table 1 of Hinshaw et al. (1996).}
\label{band_table}
\end{planotable}

\clearpage

\clearpage

\end{document}